# EVENT-BASED SEMANTIC ANALYSIS OF ACTIVITY OPPORTUNITIES: COMPARING ELDERLY RESIDENTS ACROSS CONTRASTING RESIDENTIAL CONTEXTS IN NAGOYA, JAPAN

Jianhao SHI [a], Tomio MIWA [b]
[a] *Department of Civil and Environmental Engineering, Nagoya University, Japan*
*Email: jhshi@keio.jp*
[b] *Institute of Innovation for Future Society, Nagoya University, Japan*
*Email: miwa@nagoya-u.jp*

## ABSTRACT

Population aging has made older adults' daily mobility a key issue in urban planning and transport research. This study examines whether elderly residents in different residential contexts experience unequal spatial activity opportunities in Nagoya. Using stay events derived from GPS-based mobility data, we construct a 500 m × 500 m grid-based framework to compare residents in Naka, Showa, and Moriyama across four dimensions: activity space, opportunity context, opportunity exposure, and semantic structure. The results reveal clear inter-group differences. Naka residents show the strongest concentration in the urban core and the highest opportunity exposure, Showa occupies an intermediate position, and Moriyama displays a more dispersed, corridor-oriented pattern with lower exposure. Category-specific results further show that Naka is more strongly associated with Retail/Service and Food/Drink opportunities, whereas Showa and Moriyama show relatively stronger Transit-related components. These findings indicate that spatial activity opportunity inequality is multidimensional, category-specific, and shaped by residential context.

Keywords: elderly mobility; activity space; residential context; opportunity fairness; stay events

## 1. INTRODUCTION

Population aging has made the daily mobility of older adults an increasingly important issue in urban and transport research. For older adults, daily activity spaces are closely related to accessibility, convenience, and opportunities for social participation (Levasseur et al., 2015; Zhang & Yang, 2024). Earlier studies have also shown that mobility in later life is closely tied to well-being and social inclusion rather than to movement alone (Ziegler & Schwanen, 2011; Stanley et al., 2011; He et al., 2020).

At the same time, mobility inequality cannot be understood only through movement range or spatial concentration. Activity-space research has long suggested that social and spatial differences cannot be captured adequately by residential location alone (Wong & Shaw, 2011). More recent work using mobility data has further argued that urban inequality is experienced through actual daily movement, and that observed place visits provide a basis for examining uneven place access across the city (Xu et al., 2025). GPS-based indicators have also made it possible to describe the multidimensional nature of daily mobility in a more behavior-sensitive way (Fillekes et al., 2019).

Residential context is equally important in structuring these differences. Research on housing affordability and residential mobility has shown that locational inequality can be reinforced when more vulnerable groups are sorted into less advantaged places (Baker et al., 2016). This perspective is especially relevant when comparing older adults across wards with different rent levels, because residential context may shape not only where people live but also the opportunity environments they encounter through everyday activities.

This study examines elderly residents in three wards of Nagoya—Naka, Showa, and Moriyama—which represent different rent levels and distinct urban contexts. Using stay events derived from GPS-based mobility data, the study evaluates spatial activity opportunity fairness through four dimensions: activity space, opportunity context, opportunity exposure, and semantic structure. Stay events are treated as interpretable daily activity units that link observed mobility to semantic opportunity environments (Shi et al., 2026). The paper asks whether elderly residents living in different residential settings experience unequal spatial activity opportunities in Nagoya, and whether those differences appear not only in exposure level but also in the types of opportunities embedded in daily activity spaces.

## 2. DATA AND METHODS

### 2.1 Study design and data

The study focuses on Nagoya and compares elderly residents whose inferred home wards are Naka, Showa, and Moriyama. These three wards were selected as structured comparison cases because they represent different rent levels and distinct residential contexts. Naka corresponds to a central high-opportunity setting, Showa to an intermediate mixed residential setting, and Moriyama to a more peripheral lower-opportunity setting. Figure 1 shows the study area and the spatial locations of the three wards.

The empirical basis of the analysis is a stay-event dataset derived from GPS-based mobility records. Stay events are used as interpretable daily activity units that capture where meaningful daily activities occur. The target population is older adults aged 60–69 and 70+, and all analyses are conducted by home-ward group. To represent the broader urban opportunity environment, semantic opportunity clusters are constructed from citywide POI structure. Both datasets are aggregated to a common 500 m × 500 m grid covering Nagoya.

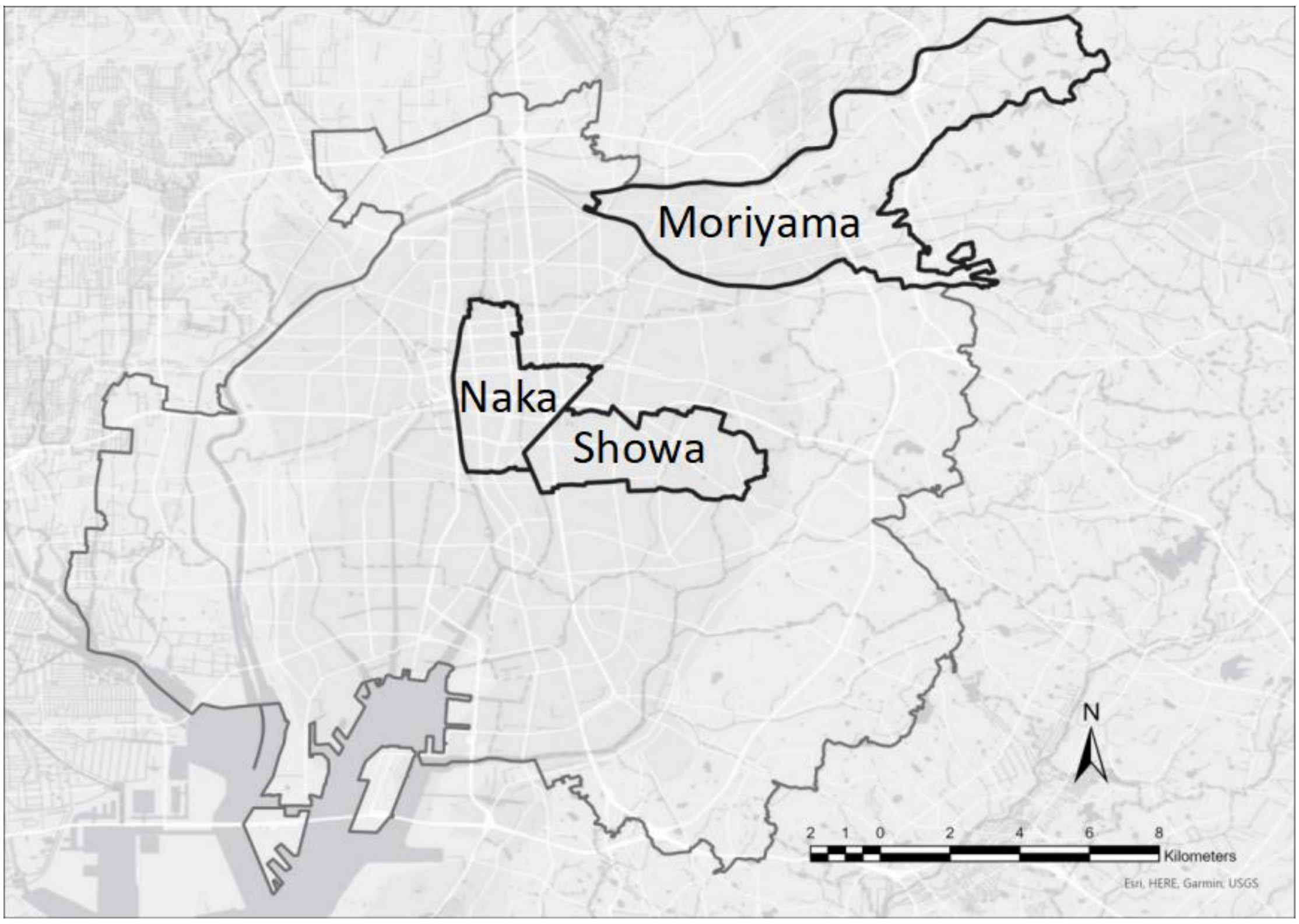

Figure 1. Study area and selected comparison wards in Nagoya
Note: Nagoya City and the boundaries of Naka, Showa, and Moriyama. These wards represent different residential contexts and rent levels.

### 2.2 Analytical framework

Figure 2 summarizes the analytical framework. The study evaluates spatial activity opportunity fairness through four dimensions: activity space, opportunity context, opportunity exposure, and semantic structure. Stay events derived from GPS data are used as interpretable daily activity units and combined with POI semantics under a structured residential comparison across the three ward groups. Figure 3 further illustrates the event-based concept, in which each stay event is understood together with the surrounding semantic opportunity environment of the grid cell where it occurs.

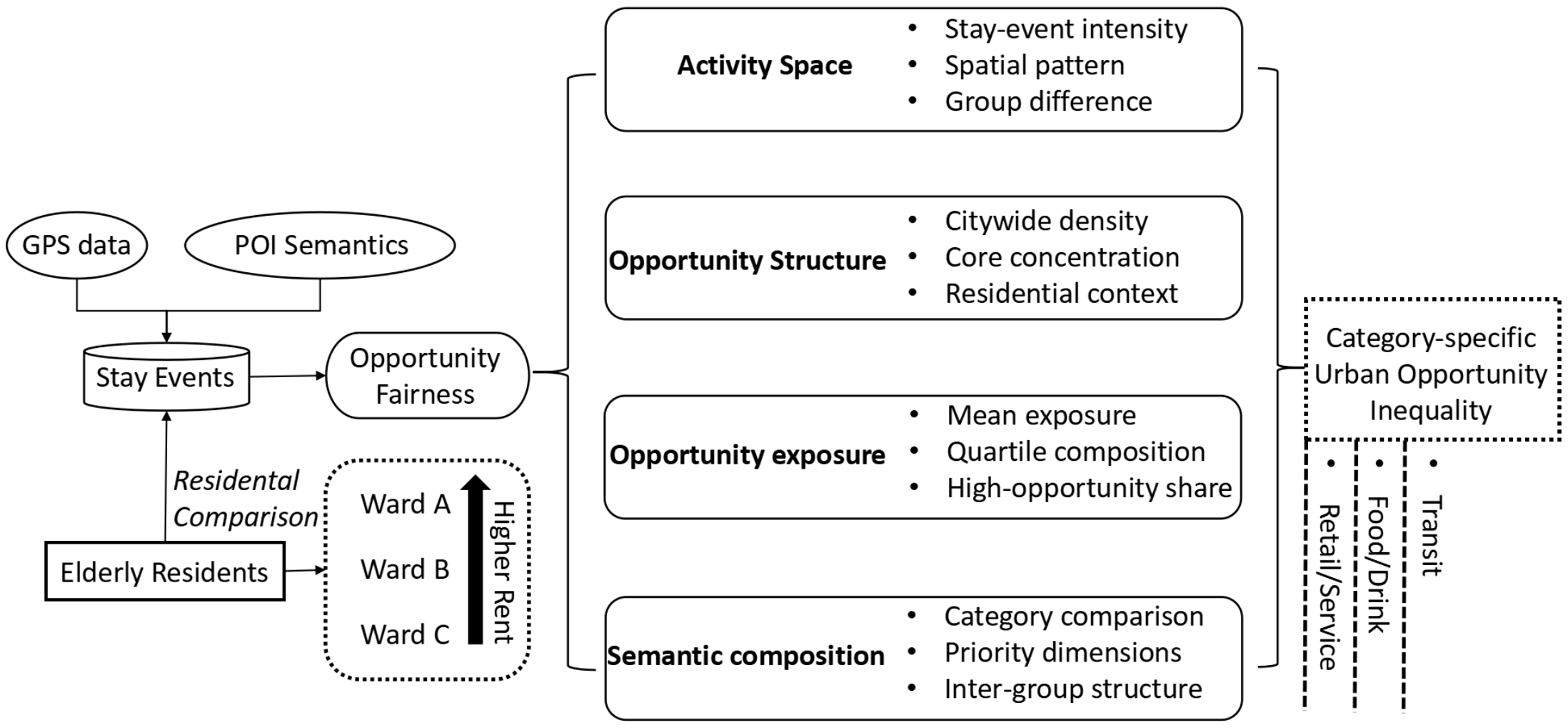

Figure 2. Analytical framework of spatial activity opportunity fairness
Note: Stay events and POI semantics are used to evaluate opportunity fairness through activity space, opportunity structure, opportunity exposure, and semantic composition.

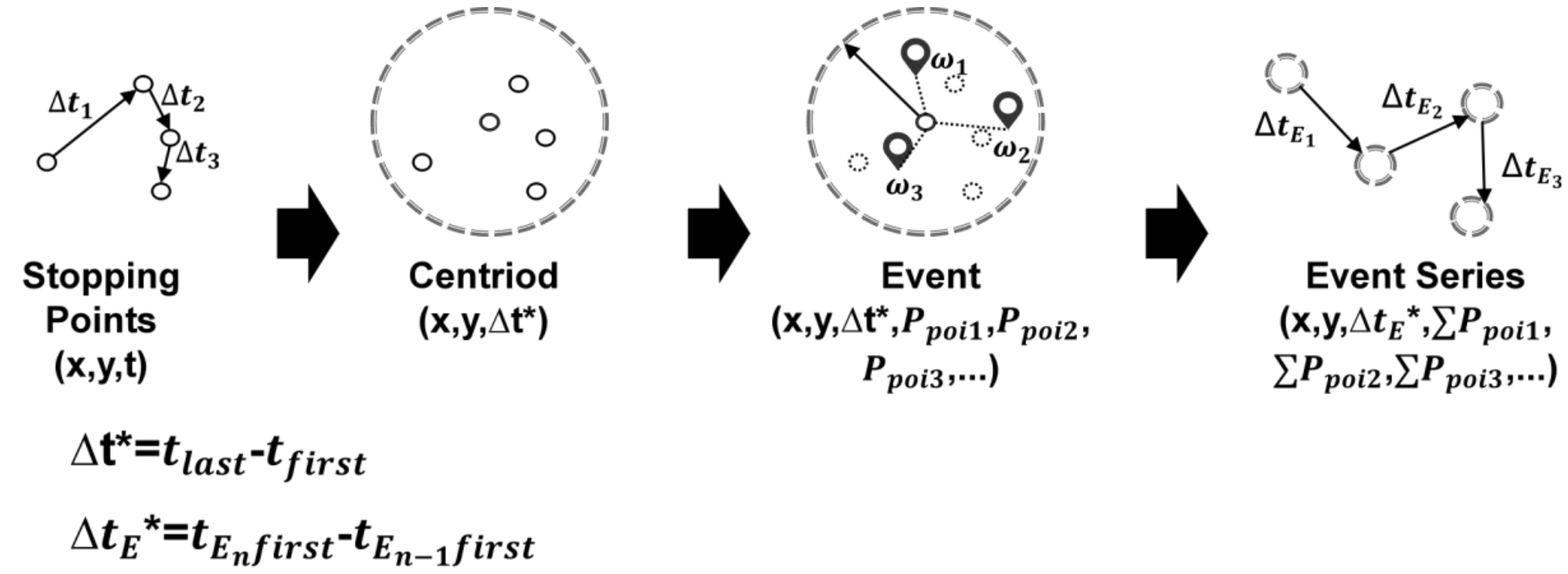

Figure 3. Conceptual illustration of stay events and surrounding opportunity environments
Note: Each stay event is treated as an interpretable daily activity unit and understood together with its surrounding semantic opportunities.

### 2.3 Measures of opportunity fairness

Opportunity fairness is examined through four measures. First, activity space is represented by stay-event density on the 500 m × 500 m grid. Second, opportunity context is represented by citywide semantic opportunity density. Third, event-level opportunity exposure is measured by assigning each stay event the semantic opportunity density of its grid cell and summarizing the result by mean exposure and quartile composition. For the quartile analysis, citywide non-zero opportunity cells are divided into four groups using cut points of 20, 36.08, and 64 clusters/km², while events in zero-opportunity cells are excluded from this classification. Fourth, semantic structure is examined through category composition and category-specific weighted intensity. Retail/Service, Food/Drink, and Transit are used as representative category-specific dimensions for mapping and comparison in the main text. In this framework, stay events are treated as interpretable daily activity units, and each event is evaluated together with the semantic opportunity environment of the grid cell in which it occurs.

## 3. RESULTS

The results show consistent differences in spatial activity opportunity fairness across the three home-ward groups. Figure 4 shows the spatial distribution of stay-event density by home-ward group and already indicates broad differences in overall activity-space organization. Building on this baseline spatial contrast, the results below focus on citywide opportunity context, event-level exposure, semantic structure, and category-specific intensity.

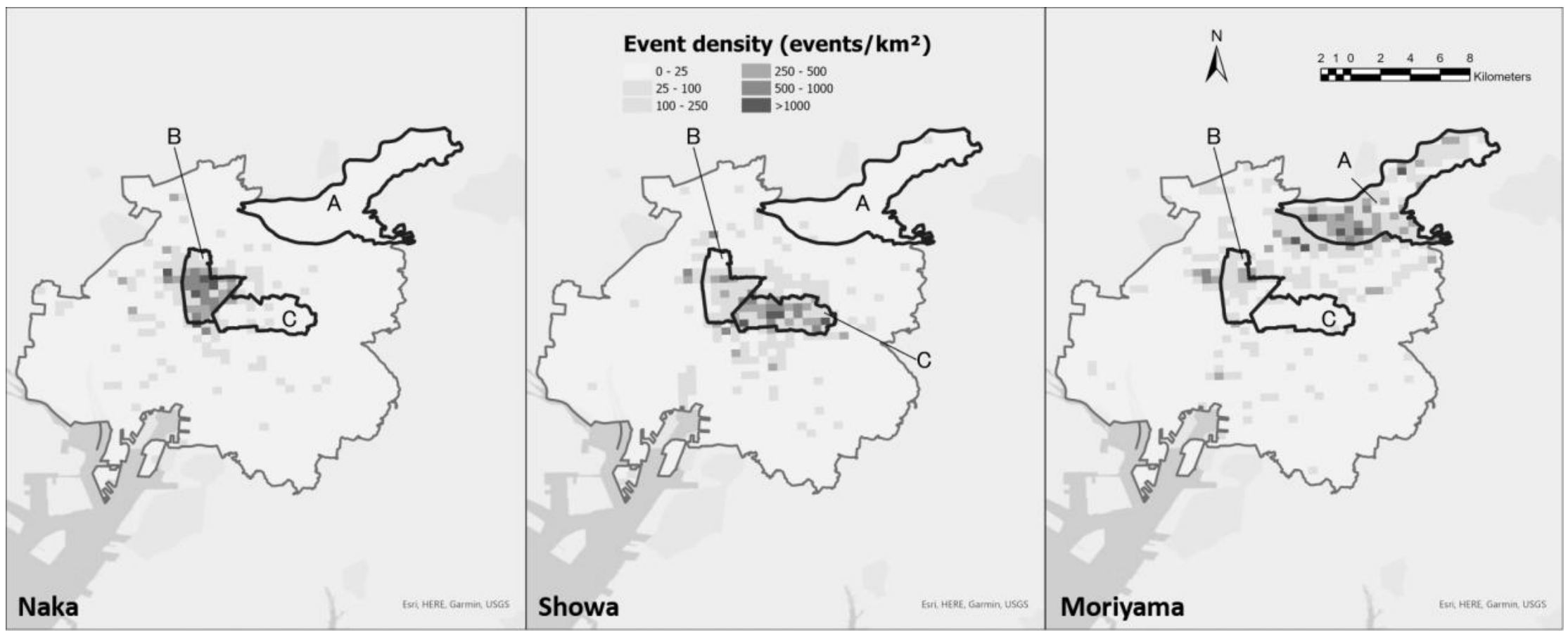

Figure 4. Spatial distribution of stay-event density by home-ward group among elderly residents in Nagoya

Note: Stay events were aggregated to a 500 m × 500 m grid and expressed as events per square kilometer. The three panels compare Naka, Showa, and Moriyama.

### 3.1 Citywide opportunity context and exposure

Figure 5 shows the citywide semantic opportunity density in Nagoya. The overall pattern is strongly uneven. High-density cells are concentrated around the urban core, while more peripheral parts of the city generally display lower opportunity density. The three selected wards are therefore embedded in different opportunity environments. Naka lies closest to the densest opportunity core, Showa occupies an intermediate position, and Moriyama is more strongly associated with a peripheral and corridor-like setting.

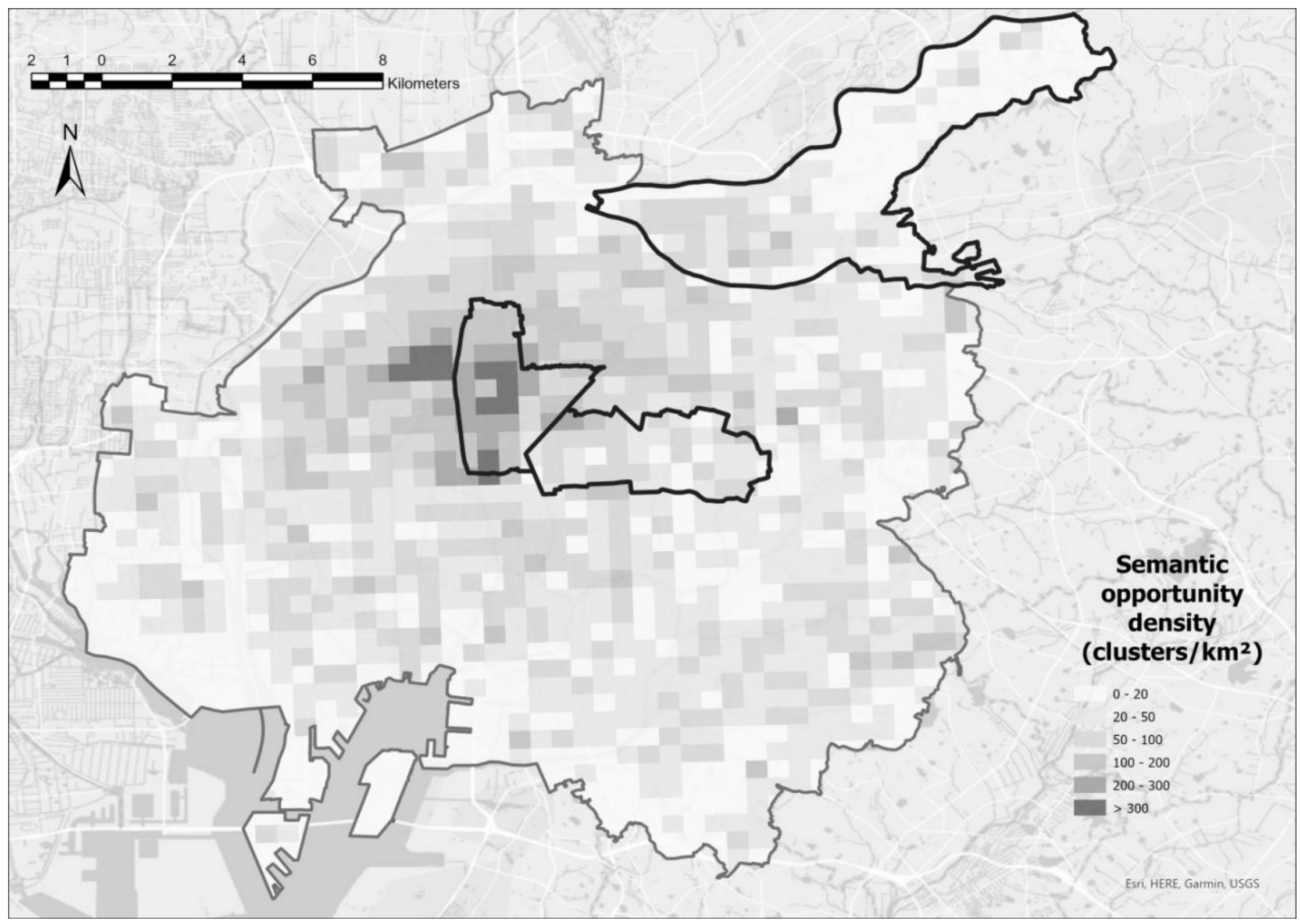

Figure 5. Citywide semantic opportunity density in Nagoya
Note: Semantic opportunity clusters were aggregated to a 500 m × 500 m grid and expressed as clusters per square kilometer. Black outlines indicate the three selected wards.

Figure 6 reports event-level opportunity exposure. Figure 6(a) shows mean opportunity exposure by ward. A clear gradient appears across the three groups, with Naka highest, followed by Showa and Moriyama. Figure 6(b) shows exposure composition across citywide quartiles. The same pattern appears more clearly. Naka has the largest share of stay events in the highest-exposure quartile, while Moriyama has the smallest. Together, these indicators show systematic differences in the level of opportunities encountered through everyday activity.

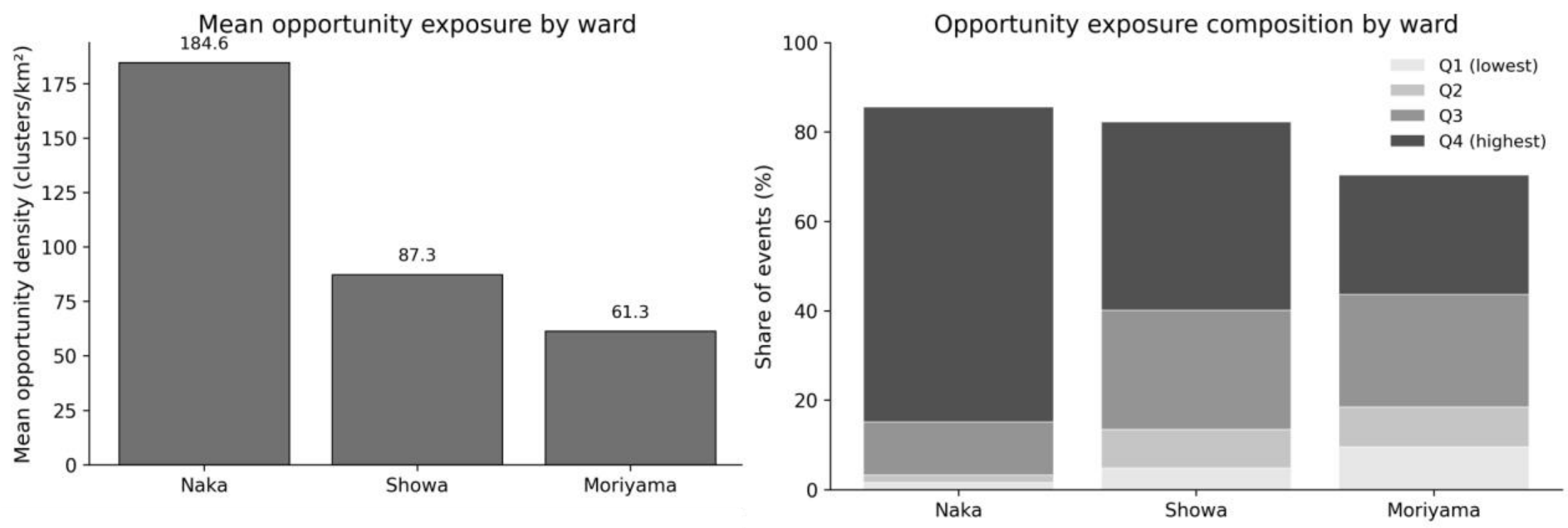

Figure 6. Opportunity exposure by home ward group (left: a; right: b)
(a)Mean opportunity exposure by ward (b) Opportunity exposure composition by ward
Note: Panel (a) shows mean opportunity exposure. Panel (b) shows the share of stay events in non-zero opportunity cells across citywide exposure quartiles.

### 3.2 Semantic structure

Figure 7 compares the semantic opportunity composition encountered by the three groups. The grouped bars show that the opportunity structures associated with daily activity differ across home-ward groups. Naka is more strongly associated with Retail/Service and Food/Drink opportunities. Showa and Moriyama show relatively stronger Transit components. This result indicates that the groups differ not only in how much opportunity they encounter, but also in what kinds of opportunities structure their daily activities.

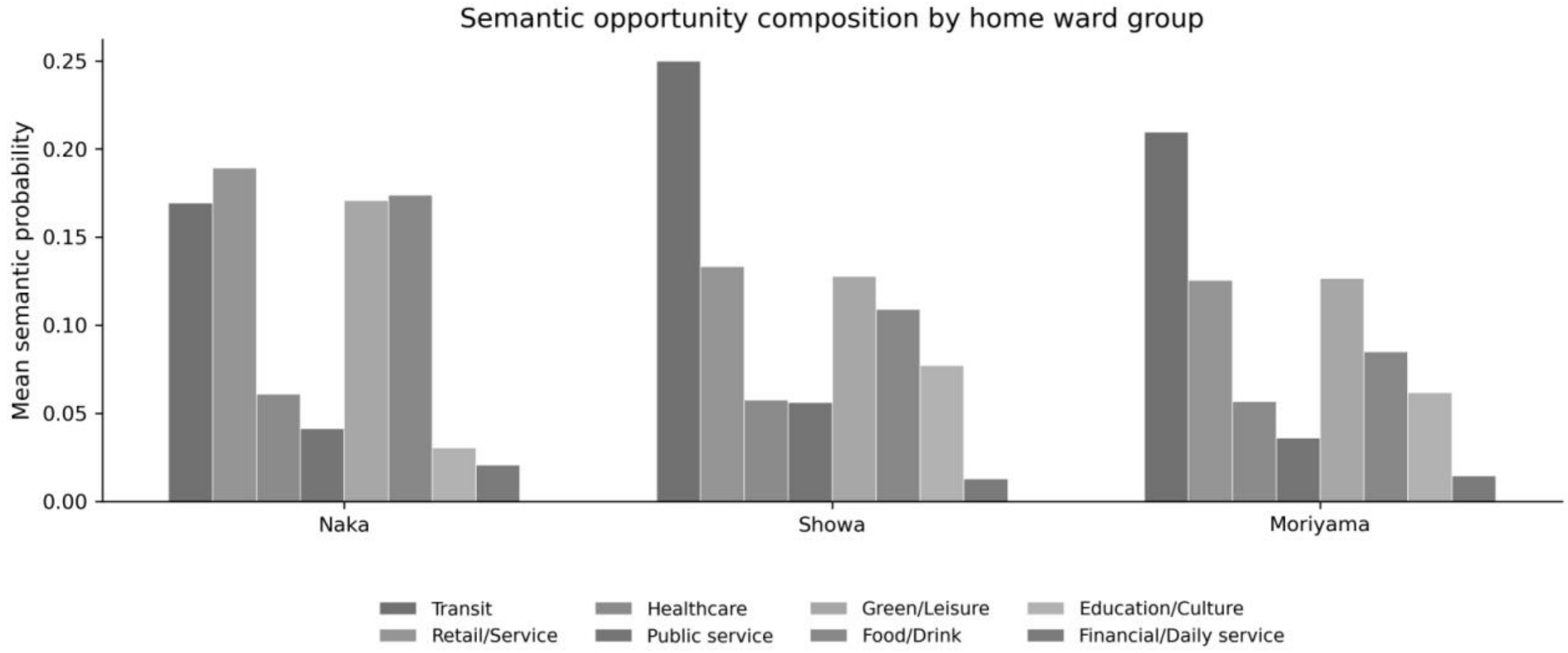

Figure 7. Semantic opportunity composition by home ward group
Note: Grouped bars show the mean semantic probability of the main opportunity categories encountered by each ward group.

### 3.3 Category-specific inequality

To examine these semantic differences more directly, the analysis focuses on Retail/Service as the clearest category-specific dimension of inequality. Figure 8 shows Retail/Service-related semantic intensity by home-ward group. Naka displays the strongest concentration in the central part of the city, indicating a high degree of embedding in service-rich environments. Showa shows a corridor-like extension from the central area, but with lower intensity than Naka. Moriyama shows a more dispersed and elongated pattern, with stronger extension toward the northeastern part of the city. The overall gradient is therefore consistent with the exposure results.

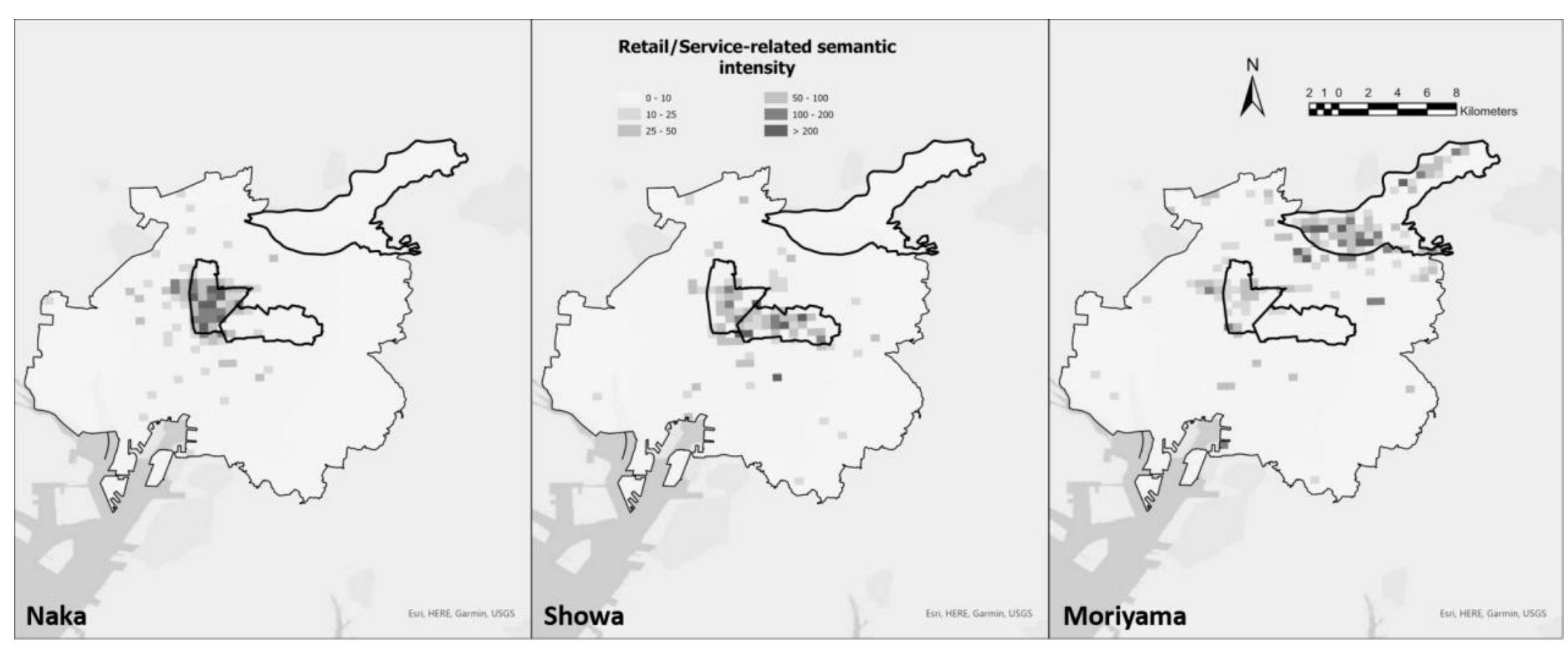

Figure 8. Retail/Service-related semantic intensity by home ward group
Note: Event-level retail/service probabilities were aggregated to a 500m × 500m grid as weighted density. All panels use the same extent, class breaks, and color scale

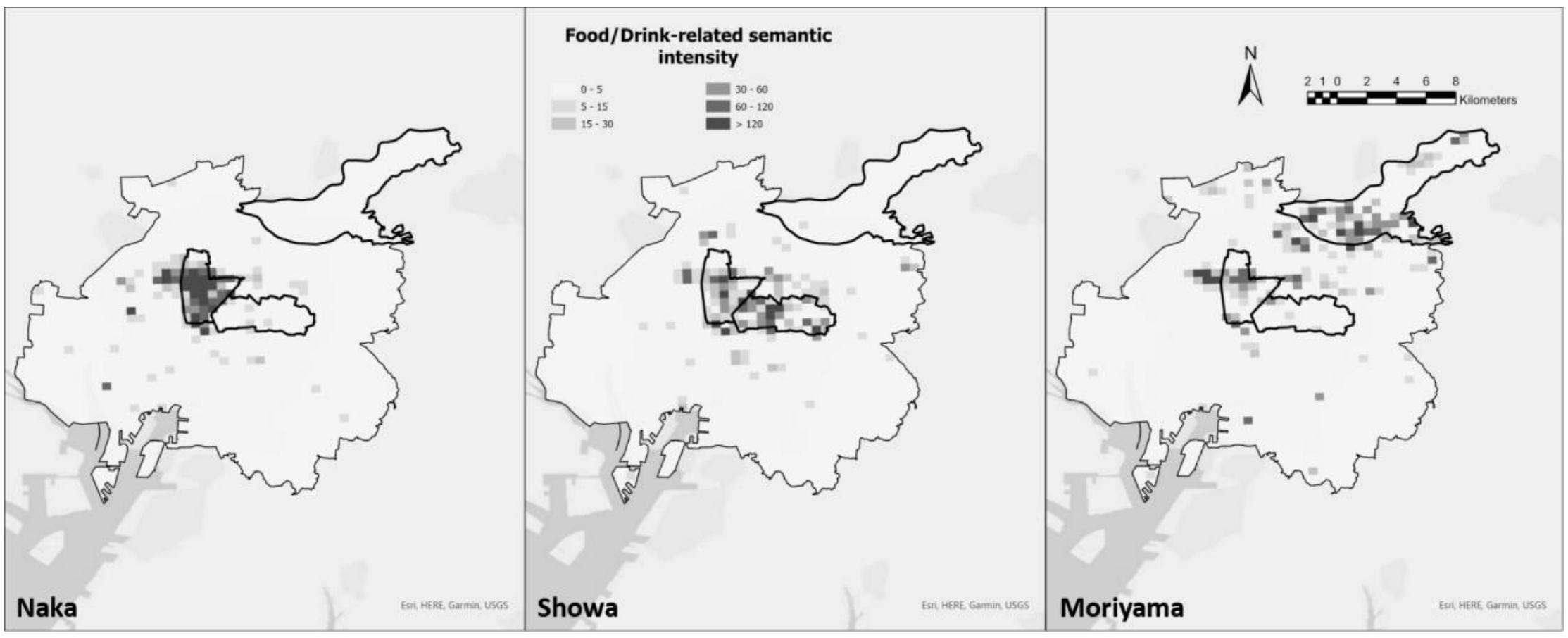

Figure 9. Food/Drink-related semantic intensity by home ward group
Note: Event-level food/drink probabilities were aggregated to a 500 m × 500 m grid as weighted density. All panels use the same extent, class breaks, and color scale.

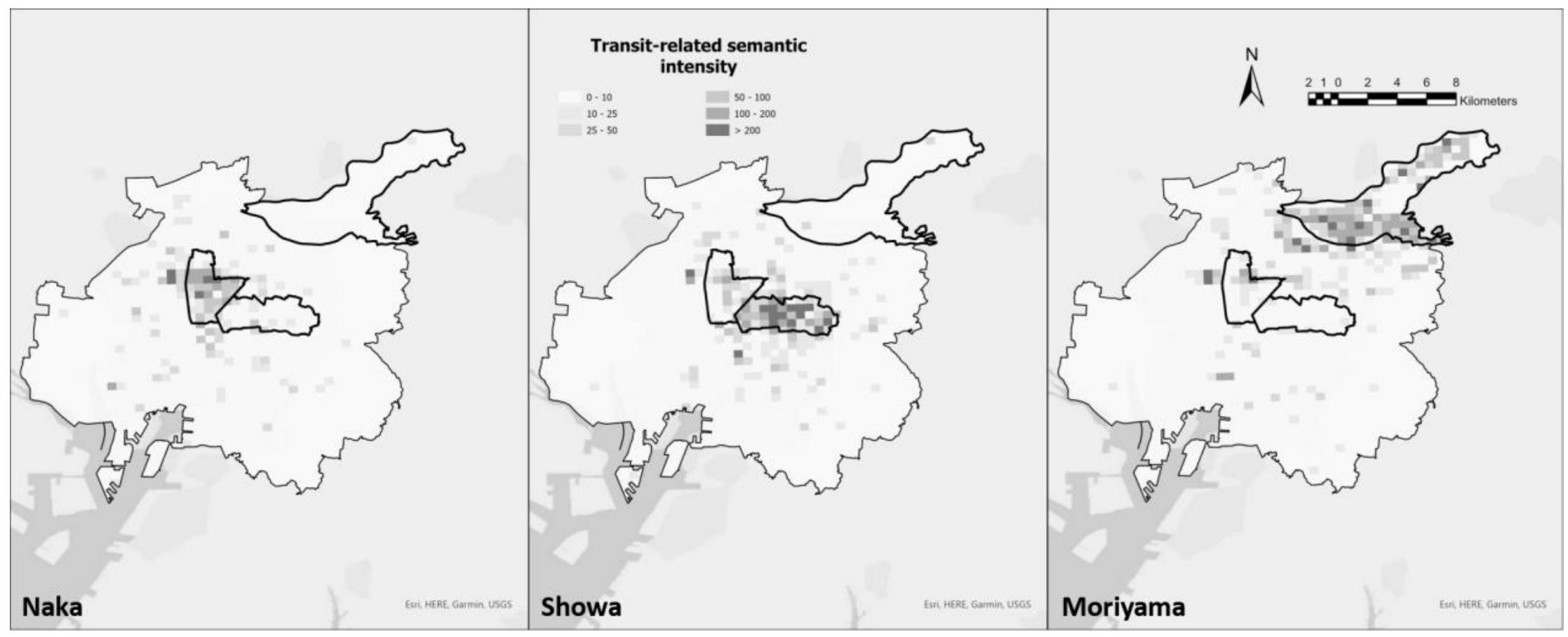

Figure 10. Transit-related semantic intensity by home ward group
Note: Event-level transit probabilities were aggregated to a 500 m × 500 m grid as weighted density. All panels use the same extent, class breaks, and color scale.

Food/Drink and Transit show related but distinct category-specific patterns and are presented in Figures 9 and 10. Food/Drink broadly follows the same direction as Retail/Service, with Naka showing stronger central embedding and higher intensity. Transit, by contrast, reflects a more corridor-based logic, especially for Showa and Moriyama, indicating that mobility-related opportunity structures differ from consumption-oriented opportunity structures. Taken together, these results indicate that spatial activity opportunity inequality among elderly residents is category-specific rather than reducible to a single center–periphery pattern.

## 4. CONCLUSION

This study compared elderly residents in three wards of Nagoya with different rent levels and examined whether they experience unequal spatial activity opportunities. The results show that citywide semantic opportunities are concentrated around the urban core and that elderly residents from different home wards encounter different levels and types of opportunities in their daily activities. Naka residents are more strongly embedded in service-rich opportunity environments, while Showa and Moriyama are associated with lower exposure and more corridor-based structures. These findings indicate that spatial activity opportunity inequality among elderly residents is multi-dimensional and shaped by residential context. Future work should extend the analysis to the full city, include multiple age groups, and examine temporal inequality through stay-event duration.

## 5. REFERENCES

Baker, E., Bentley, R., Lester, L. and Beer, A. (2016) Housing affordability and residential mobility as drivers of locational inequality. *Applied Geography*, 72, pp. 65–75.

Fillekes, M.P., Giannouli, E., Kim, E.-K., Zijlstra, W. and Weibel, R. (2019) Towards a comprehensive set of GPS-based indicators reflecting the multidimensional nature of daily mobility for applications in health and aging research. *International Journal of Health Geographics*, 18, Article 17.

He, S.Y., Thøgersen, J., Cheung, Y.H.Y. and Yu, A.H.Y. (2020) Ageing in a transit-oriented city: Satisfaction with transport, social inclusion and wellbeing. *Transport Policy*, 97, pp. 85–94.

Levasseur, M., Généreux, M., Bruneau, J.-F., Vanasse, A., Chabot, É., Beaulac, C. and Bédard, M.-M. (2015) Importance of proximity to resources, social support, transportation and neighborhood security for mobility and social participation in older adults: Results from a scoping study. *BMC Public Health*, 15, Article 503.

Shi, J., Miwa, T. and Yan, W. (2026) Rhythm-consistent semi-Markov simulation of tourist mobility rhythms with probabilistic event-to-POI assignment. *arXiv preprint*. Available at: https://doi.org/10.48550/arXiv.2604.06672

Stanley, J.K., Hensher, D.A., Stanley, J.R. and Vella-Brodrick, D. (2011) Mobility, social exclusion and well-being: Exploring the links. *Transportation Research Part A: Policy and Practice*, 45(8), pp. 789–801.

Wong, D.W.S. and Shaw, S.-L. (2011) Measuring segregation: An activity space approach. *Journal of Geographical Systems*, 13(2), pp. 127–145.

Xu, F., Wang, Q., Moro, E., Chen, L., Salazar Miranda, A., González, M.C., Tizzoni, M., Song, C., Ratti, C., Bettencourt, L., Li, Y. and Evans, J. (2025) Using human mobility data to quantify experienced urban inequalities. *Nature Human Behaviour*, 9(4), pp. 654–664.

Zhang, N. and Yang, Q. (2024) Public transport inclusion and active aging: A systematic review on elderly mobility. *Journal of Traffic and Transportation Engineering (English Edition)*, 11(2), pp. 312–347.

Ziegler, F. and Schwanen, T. (2011) "I like to go out to be energised by different people": An exploratory analysis of mobility and wellbeing in later life. *Ageing & Society*, 31(5), pp. 758–781.